\documentclass[twoside]{article}
\usepackage{amsmath,amsfonts,amsthm}
\usepackage[affil-it]{authblk}

\usepackage{amstext}
\usepackage{amsgen}
\usepackage{amsbsy}
\usepackage{amsopn}
\usepackage{amssymb}
\usepackage{graphicx}
\usepackage{epsfig}

\newtheorem{theorema}{Theorem}
\newtheorem{algorithm}[theorema]{Algorithm}

\newtheorem{prop}{Theorem}
\newtheorem{proposition}[prop]{Proposition}

\newcommand{\Prob}{$\mathcal{P}$}

\newcommand{\SSAT}{{SSAT}$(n,m)$}

\def\IMAGESPATH{.}

\begin{document}
\hyphenation{corres-pond}

\title{Lower bound for the Complexity of the Boolean Satisfiability Problem}

\author{Carlos Barr\'{o}n-Romero
\thanks{Universidad Aut\'{o}noma Metropolitana, Unidad
 Azcapotzalco, Av. San Pablo No. 180, Col. Reynosa Tamaulipas,
 C.P. 02200. MEXICO.
%\email{ cbarron@correo.azc.uam.mx}
 }
}
\date{February 12, 2015}

\maketitle

\begin{abstract}

This paper depicts algorithms for solving  the decision Boolean
Satisfiability Problem. An extreme problem is formulated to
analyze the complexity of algorithms and the complexity for
solving it. A novel and easy reformulation as a lottery for an
extreme case is presented to determine a stable complexity around
$2^n$. The reformulation point out that the decision Boolean
Satisfiability Problem can only be solved in exponential time.
This implies there is not an efficient algorithm for the NP Class.

\end{abstract}

% \begin{keywords}
% \textbf{Key words.}:
 Algorithms, Complexity, SAT, NP,
Quantum Computation.
% \bigskip
% \end{keywords}

% \begin{AMS}
% \textbf{AMS subject classifications.}
68Q10, 68Q12,68Q19,68Q25.
% \end{AMS}

\pagestyle{myheadings} \thispagestyle{plain} \markboth{Carlos
 Barr\'{o}n-Romero}{Lower bound for SSAT}

%%%%%%%%%%%%%%%%%%%%%%%%%%%%%%%%%%%%%%%%%%%%%%%%%%%%%%%%%%%%%%%%%%%%%%%%%%%%
%%%%%%%%%%%%%%%%%%%%%%%%%%%%%%%%%%%%%%%%%%%%%%%%%%%%%%%%%%%%%%%%%%%%%%%%%%%%
\section{Introduction}
%%%%%%%%%%%%%%%%%%%%%%%%%%%%%%%%%%%%%%%%%%%%%%%%%%%%%%%%%%%%%%%%%%%%%%%%%%%%
%%%%%%%%%%%%%%%%%%%%%%%%%%%%%%%%%%%%%%%%%%%%%%%%%%%%%%%%%%%%%%%%%%%%%%%%%%%%
%%%%%%%%%%%%%%%%%%%%%%%%%%%%%%%%%%%%%%%%%%%%%%%%%%%%%%%%%%%%%%%%%%%%%%%%%%%%
%%%%%%%%%%%%%%%%%%%%%%%%%%%%%%%%%%%%%%%%%%%%%%%%%%%%%%%%%%%%%%%%%%%%%%%%%%%%

My previous works over the NP class is~\cite{arXiv:Barron2005},
~\cite{arXiv:Barron2010}, and ~\cite{arXiv:Barron2015b}. In the
last one, the classical decision problem, the Boolean
Satisfiability Problem, named SAT was used to state a lower bound
for its complexity.

As a general framework, my technique consists: 1) to study general
problem, 2) to determine a simple reduction, and 3) to analyze for
trying to build an efficient algorithm for the simple problem. I
like to explains that to build an algorithms to determine a
complexity bound has more than I depicts
in~\cite{arXiv:Barron2015b}. I take the approach by similarities
from applied mathematics: the well-know optimization conditions,
the search inside of a region or outside of it, and fixed point
method. My article \cite{arXiv:Barron2015b} focus on describing
the fixed point and probabilistic approach. This article is  a
commentary study of the decision SAT, the changes in the
algorithms presented here do not change my main result for NP's
complexity but they clarifies details.

This paper focus in the  the SAT's properties, objections and
proofs about the algorithms for solving an extreme case problem of
SAT (also I called reduced SAT or Simple SAT, see
section~\ref{sc:SAS_SSAT}.). Hereafter, SSAT states Simple SAT.
The following section depicts SAT and SSAT and their properties.
The next section depicts the algorithms for SSAT, and the
complexity for an extreme SSAT is depicted in the next section.
Some parts of \cite{arXiv:Barron2015b} are repeated here to make
this article self-content.

%%%%%%%%%%%%%%%%%%%%%%%%%%%%%%%%%%%%%%%%%%%%%%%%%%%%%%%%%%%%%%%%%%%%%%%%%%%%
%%%%%%%%%%%%%%%%%%%%%%%%%%%%%%%%%%%%%%%%%%%%%%%%%%%%%%%%%%%%%%%%%%%%%%%%%%%%
\section{SAT and Simple SAT}~\label{sc:SAS_SSAT}
%%%%%%%%%%%%%%%%%%%%%%%%%%%%%%%%%%%%%%%%%%%%%%%%%%%%%%%%%%%%%%%%%%%%%%%%%%%%
%%%%%%%%%%%%%%%%%%%%%%%%%%%%%%%%%%%%%%%%%%%%%%%%%%%%%%%%%%%%%%%%%%%%%%%%%%%%
%%%%%%%%%%%%%%%%%%%%%%%%%%%%%%%%%%%%%%%%%%%%%%%%%%%%%%%%%%%%%%%%%%%%%%%%%%%%
%%%%%%%%%%%%%%%%%%%%%%%%%%%%%%%%%%%%%%%%%%%%%%%%%%%%%%%%%%%%%%%%%%%%%%%%%%%%

A Boolean variable only takes the values: $0$ (false) or $1$
(true). The logical operators are \textbf{not}: $\overline{x};$
\textbf{and}: $\wedge ,$ and \textbf{or}: $\vee .$

 Hereafter, $\Sigma =\left\{0,1\right\}$ is the corresponding alphabet, $x$ is a binary string in
$\Sigma^n$ means its corresponding number in $[0,2^n-1]$ and
reciprocally. The inner or fixed point approach means to take the
data from the translation of the problem's formulas, and outside
or probabilistic approach means to take randomly the data from the
problem's search space.

A SAT$(n,m)$ problem consists to answer if a system of $m$ Boolean
formulas in conjunctive normal form over $n$ Boolean variables has
an assignation of logical values such the system of formulas are
true.

The system of formulas is represented as a matrix, where each row
correspond to a disjunctive formula. By example, let SAT$(4,4)$ be

\begin{equation*}
\begin{array}{ccccc}
\ \    & (x_{3}& \vee \ \overline{x}_{2}& & \vee \  x_{0})  \\
\wedge &   &  ( x_{2}& \vee \ x_{1} & \vee \  x_{0})\\
\wedge &  \         &  ( \overline{x}_{2} & \vee \  x_{1} & \vee \
x_{0}) \\
\wedge &(x_3 & & & \ \vee \ \overline{x}_{0}).%
\end{array}%
\end{equation*}

This problem is satisfactory. The assignation
$x_{0}=1,x_{1}=0,x_{2}=1,$ and $x_{3}=1$ is a solution, as it is
depicting by substituting the Boolean values:
\begin{equation*}
\begin{array}{ccccc}
\ \    & (1& \vee \ 0 & & \vee \  1) \\
\wedge &  & ( 1 & \vee \ 0  &  \vee\  1)\\
\wedge &          &  ( 0 & \vee \  0 & \vee \
1) \\
\wedge & (1  & & & \vee \ 0) %
\end{array}%
\equiv 1.
\end{equation*}

It is important to note that the requirement of rows with the same
number of Boolean variables in a given order is a simple reduction
for studying SAT. This paper focuses in  this simple formulation
of SAT. SSAT$(n,m)$ is a SAT where its $m$ Boolean row formulas
have the same length and the Boolean variables are in each row are
in the same order, $x_{n-1},\ldots,x_0$.

For any SSAT, each row of the system of Boolean formulas can be
translated into a set of binary numbers.

Each row of SSAT maps to a binary string in $\Sigma^n$, with the
convention: $\overline{x}_{i}$ to $0$ (false), and $x_{i}$ to $1$
(true) in the $i$ position. Hereafter, any binary string in
$\Sigma^n$ represents a binary number and reciprocally.

For example, given SAT$(2,2)$:
\begin{equation*}
\begin{array}{ccc}
\ \ & (  \overline{x}_{1} & \vee \  x_{0}) \\
\wedge & ( x_{1} & \vee  \  x_{0} ).%
\end{array}%
\end{equation*}

It is traduced to:
\begin{equation*}
\begin{array}{l}
01 \\
11.%
\end{array}%
\end{equation*}%

The problem is to determine, does SSAT$(n,m)$ have a solution?
without previous knowledge.

%%%%%%%%%%%%%%%%%%%%%%%%%%%%%%%%%%%%%%%%%%%%%%%%%%%%%%%%%%%%%%%%%%%%%%%%%%%%
%%%%%%%%%%%%%%%%%%%%%%%%%%%%%%%%%%%%%%%%%%%%%%%%%%%%%%%%%%%%%%%%%%%%%%%%%%%%
\section{Characteristics and properties of SSAT}~\label{sc:CharPrpSSAT}

\begin{proposition}
~\label{prop:SATvsSSAT} 1) A problem SAT can be transformed into
an equivalent SSAT. 2) A problem SSAT  is a SAT. 3) SSAT could be
a subproblem of a problem SAT.
\begin{proof}
1) A SAT is transformed into an equivalent SSAT by algebraic
procedures based in $F \equiv F  \wedge (v \vee \overline{v})$,
where $F$ is a formula and $v$ is a Boolean variable. 2) Any SSAT
is a SAT with formulas of the same number of variables 3) On the
other hand, SAT could have a subset of the Boolean formulas, that
they can be arranged as a SSAT.
\end{proof}
\end{proposition}

For the cases 2 and 3) the complexity for solving SSAT is less
than the complexity for solving SAT. The case 1) opens the
possibility that for  some SAT can be solved with less complexity
than solving SSAT. By example, SAT$(2,2)$ for $x_1,x_0$ under
$(x_0) \wedge (\overline{x}_0) \equiv 0$ versus SSAT$(2,4)$
$(\overline{x}_1 \vee \overline{x}_0) \wedge (\overline{x}_1 \vee
x_0) \wedge (x_1 \vee \overline{x}_0)\wedge (x_1 \vee x_0) \equiv
0.$ However, the first system can be see as the SSAT$(1,2)$ $(x_0)
\wedge (\overline{x}_0)$, which has no solution. This article
focuses in study SSAT, in my next article, the complexity SSAT
$\preceq$ SAT is depicted in detail.

\begin{proposition}
~\label{prop:TradSAT}
\begin{enumerate}
    \item Any SAT$(n,m)$ can be translated to a matrix of ternary numbers,
and the ternary numbers are strings in $\{0,1,2\}^n$.
    \item The search
space of SSAT is less than the search space of SAT.
\end{enumerate}
\begin{proof}
\begin{enumerate}
    \item Taking the alphabet $\left\{ 0,1,2\right\}.$ Each row of SAT is mapping to a ternary number,
    with the convention: $\overline{x}_{i}$ to
$0$ (false), $x_{i}$ to $1$ ( true), and $2$ when the variable
$x_{i}$ is no present.
    \item By construction, $|\Sigma^n|$ $=$ $|\left\{ 0,1\right\}^n|$ $=$ $2^n$
    $\leq$ $3^n$ $=$ $|\left\{ 0,1,2\right\}^n|.$
\end{enumerate}
\end{proof}
\end{proposition}

The previous propositions justify to focus in SSAT. The former
proposition states that sections of a SAT can be see as subproblem
type SSAT. Moreover, it is sufficient to prove that there is not
polynomial time algorithm for it.

By example, the previous SAT$(4,4),$ it contains the following
SSAT$(3,2):$
\begin{equation*}
\begin{array}{cccc}
& (\ x_{2}& \vee \ x_{1} & \vee \  x_0)  \\
\wedge & (\ \overline{x}_{2} & \vee \  x_{1} & \vee \  x_0).%
\end{array}%
\end{equation*}

\begin{proposition}
~\label{prop:binNumBlock} Given a binary number $b$ $=$
$b_{n-1}b_n\ldots b_0$. Then the Boolean disjunctive formula that
correspond to the translation of $\overline{b}$ is 0.
\begin{proof}
Without loss of generality, let $x$ be $=$ $x_{n-1}\vee
\overline{x}_{n-2}\vee \ldots \vee \overline{x}_0$ the translation
of $b$ and $\overline{x}$ $=$ $\overline{x}_{n-1}\vee x_{n-2}\vee
\ldots \vee x_0$ the translation of $\overline{b}$. Then $x \wedge
\overline{x}$ $=$ $(x_{n-1}\vee \overline{x}_{n-2}\vee \ldots \vee
\overline{x}_0)\wedge(\overline{x}_{n-1}\vee x_{n-2}\vee \ldots
\vee x_0)$ $=$ $(x_{n-1}\wedge \overline{x}_{n-1}) \vee
(\overline{x}_{n-2} \wedge x_{n-2}) \vee \ldots \vee
(\overline{x}_0 \wedge x_0)$ $=$ $0$.
\end{proof}
\end{proposition}

The translation of the rows formulas of SSAT allows to define a
table of binary numbers for SSAT. The matrix of binary values is
an equivalent visual formulation of SSAT$(n,m)$. The following
boards have not a set of values in $\Sigma $ to satisfy them:

\begin{equation*}
\begin{tabular}{|c|}
\hline $x_{1}$ \\ \hline 1 \\ \hline 0 \\
\hline
\end{tabular}%
\ \ \
\begin{tabular}{|l|l|}
\hline $x_{2}$ & $x_{1}$ \\ \hline 0 & 0 \\ \hline
1 & 1 \\ \hline 0 & 1  \\
\hline 1 & 0 \\ \hline
\end{tabular}%
\end{equation*}

I called unsatisfactory boards to the previous ones. It is clear
that they have not a solution because each binary number has its
binary complement. To find an unsatisfactory board is like order
the number and its complement, by example: $000,$ $101,$ $110,$
$001,$ $010,$ $111,$ $011,$ and $100$ correspond to the
unsatisfactory board, i.e.:
\begin{equation*}
\begin{array}{c}
000 \\
111 \\
001 \\
110 \\
010 \\
101 \\
011 \\
100.%
\end{array}%
\end{equation*}

By inspection, it is possible to verify that the previous binary
numbers correspond to a SSAT$(3,8)$ with no solution because any
binary number is blocked by its complement binary number (see
prop.~\ref{prop:binNumBlock}). By example,   $000$ and $111$
correspond to $(x_2 \vee x_1 \vee x_0) \wedge (\overline{x}_2 \vee
\overline{x}_1 \vee \overline{x}_0)$. Substituting by example
$x_2=1,x_1=1,x_0=1$, we get $(1 \vee 1 \vee 1) \wedge (0 \vee 0
\vee 0)$ $\equiv$ $(1) \wedge (0)$ $\equiv$ $0.$

\begin{proposition}
~\label{prop:SolSAT_binary} SSAT$(n,m)$ has different rows
 and $m<2^{n}$. There is a satisfactory assignation that
correspond to a binary string in $\Sigma^n$ as a number from $0$
to $2^{n}-1$.
\begin{proof}
Let $s$ be any binary string that corresponds to a binary number
from $0$ to $2^n-1$, where $s$  has not its complement into the
translated formulas of the given SSAT$(n,m)$. Then $s$ coincide
with at least one binary digit of each binary number of the
translated rows formulas, the corresponding Boolean variable is 1.
Therefore, all rows are 1, i.e., $s$ makes SSAT$(n,m)$ = 1.
\end{proof}
\end{proposition}

The previous proposition point out when a solution $s\in[0,
2^n-1]$ exists for SSAT. More important, SSAT can be see like the
problem to look for a number $s$ which its complements does not
corresponded to the translated numbers of the SSAT's formulas.

\begin{proposition}
~\label{prop:NoSolSAT_binary} SSAT$(n,2^n)$'s rows correspond to
the $0$ to $2^{n}-1$ binary numbers. Then it is an unsatisfactory
board.
\begin{proof}
The binary strings of the values from $0$ to $2^n-1$ are all
possible assignation of values for the board. These strings
correspond to all combinations of $\Sigma^n$, and by the
prop.~\ref{prop:binNumBlock} SSAT$(n,2^n)$ has not solution.
\end{proof}
\end{proposition}

This proposition~\ref{prop:NoSolSAT_binary} states that if $m=2^n$
and SSAT has different rows, then there is not a solution. These
are necessary conditions for any SSAT but these conditions a)
different rows formulas and b) the number of rows formulas are
previous knowledge.

As it is depicted below, it is possible to evaluate SSAT$(n,m)$ as
a logic circuit without substituting, and evaluating the Boolean
formulas, i.e., without knowing the rows of SSAT.

\begin{proposition}
~\label{prop:NoSolSAT} Given SAT$(n,m).$ There is not solution, if
$L$ exists, where $L$ is any subset of Boolean variables, with
their rows formulas isomorphic to an unsatisfactory board.
\begin{proof}
The subset $L$ satisfies the
proposition~\ref{prop:NoSolSAT_binary}. Therefore, it is not
possible to find satisfactory set of $n$ values for SAT$(n,m)$.
\end{proof}
\end{proposition}

Here, the last proposition depicts a necessary condition in order
to determine the existence of the solution for SSAT. It is easy to
understand but it is quite different to accept that SSAT has not
solution, i.e., that SSAT is equivalent to an unsatisfactory
board. The next propositions justifies focus in an extreme SSAT because solving some easy cases of SSAT can be solved in very efficient time without a satisfactory assignation as a witness.

\begin{proposition}
~\label{prop:O1SSAT} Given SSAT$(n,m).$
If $m < 2^n$ then SSAT$(n,m)$ has a solution, such answer is found with complexity  $\mathbf{O}(1)$.
\begin{proof}
The rows of the given SSAT$(n,m)$ do not correspond to all numbers in the search space $[0,2^n-1],$ even with repeated rows. Then, it exits a number which is not blocked.

The complexity is $\mathbf{O}(1)$, the only step corresponds to "if $m < 2^n$ then SSAT$(n,m)$ has a solution".
\end{proof}
\end{proposition}

\begin{proposition}
~\label{prop:OKSSAT} Given SSAT$(n,m).$ Let $k$ be the number of failed candidates of the search space $[0,2^n]$, such $k=k_1+k_2$ where $k_1$ is the number of candidates that their translation is a formula of SSAT$(n,m)$, and  $k_2$ is the number of candidates that their translation is a repeated formula in SSAT$(n,m).$

If $m-k_2 < 2^n$ then SSAT$(n,m)$ has a solution, such answer is found with complexity  $\mathbf{O}(k)$.
\begin{proof}
$k_1+ m-k$ is an estimation of the rows of the given SSAT$(n,m)$. They do not correspond to all numbers in the search space $[0,2^n-1],$ because $k_1+ m-k$ $=$ $k_1+ m-(k_1 + k_2)$ $=$ $m - k_2$ $< 2^n$. Then, SSAT$(n,m)$ has a satisfactory assignation, i.e., it is not a blocked board.

The complexity is $\mathbf{O}(k)$. It corresponds to the $k$ tested candidates.
\end{proof}
\end{proposition}

The previous propositions do not estimate a witness $x$ for verifying that SSAT$(n,m)(x) = 1.$ SSAT$(n,m)$ has a satisfactory assignation is implied by the fact that such SSAT$(n,m)$ is not a blocked board.

\begin{figure}[tbp]
\centerline{\psfig{figure=\IMAGESPATH/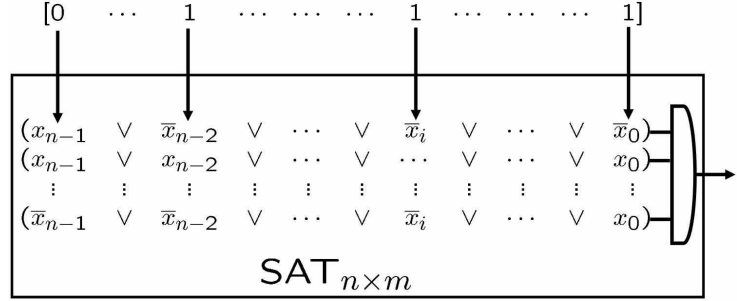, height=40mm}}~
\caption{ SAT$(n,m)$ is a white of box containing a circuit of
logical gates where each row has the same number of Boolean
variables.} \label{fig:BoxSATnxm}
\end{figure}

SSAT can be see as a logic circuit, it only depends of the
selection of the  binary values assigned to $n$ lines, each line
inputs the corresponding binary value  to its Boolean variable
$x_i$. This is an important consideration because the complexity
of the evaluation as a logic function is $\mathbf{O}(1)$. The
figure~\ref{fig:BoxSATnxm} depicts SAT$(n,m)$ as its logic
circuit.

\begin{proposition}
~\label{prop:SAT_TwoOne} Given SSAT$(n,m)$ as a circuit, and $M_{n \times m}$ the numbers of the translation of the SSAT$(n,m)$'s rows.
\begin{enumerate}
    \item Let $k$ be the translation of any row formula of SSAT$(n,m)$.
    \item Let $k$ be any binary number, $k$ $\in$ $[0: 2^n-1]$.
\end{enumerate}
if SSAT${(n,m)}$$(k)=0$, then
\begin{enumerate}
    \item SSAT$(n,m)$ $(\overline{k})=0$ and $k, \overline{k}\in M_{n \times m}$.
    \item SSAT${(n,m)}$ $(k)=0$ and $\overline{k}\in M_{n \times m}$.
\end{enumerate}

\begin{proof}
Without previous knowledge in the second case, the information
that we have is SSAT${(n,m)}$ $(k)=0$. It is caused by the translation of $\overline{k}$ in SSAT$(n,m).$ On the former case when
SSAT${(n,m)}$ $(k)=0$ is not satisfied, it is because the
complement of $k$ blocks the  system, i.e. SSAT${(n,m)}$
$(\overline{k})=0$ (see prop.~\ref{prop:binNumBlock}) and $k, \overline{k}\in M_{n \times m}$.
\end{proof}
\end{proposition}

\begin{proposition}
$\Sigma={0,1}$ is an alphabet. Given SSAT$(n,m)$, the set
$\mathcal{S}$ $=$ $\{ x \in \Sigma^n |$ SSAT$(n,m)$$(x)$$=1$ $\}$
$\subset \Sigma^n$ of the satisfactory assignations is a regular
expression.
\begin{proof}
$\mathcal{S} \subset \Sigma^n$.
\end{proof}
\end{proposition}

The last proposition depicts that a set of binary strings
$\mathcal{S}$ of the satisfactory assignations can be computed by
testing SSAT$(n,m)$$(x)$$=1$, and the cost to determine
$\mathcal{S}$ is $2^n$, the number of different strings in
$\Sigma^n$.

With $\mathcal{S}$ $\neq$ $\emptyset$ there is not opposition to
accept that SSAT$(n,m)$ has solution, no matters if $m$ is huge
and the formulas are in disorder or repeated. It is enough and
sufficient to evaluate SSAT$(n,m)$$(x)$, $x\in \mathcal{S}$.

On the other hand, $\mathcal{S}$ $=$ $\emptyset$, there is not a
direct verification. It is necessary, to validate how
$\mathcal{S}$ is constructed. Solving \SSAT  could be easy if we
have the binary numbers that has not a complement in its
translated rows. Also, because, $|\Sigma^n|=2^n$ has exponential
size, it could not be convenient to focus in the information of
\SSAT \ with $m\gg 2^n$.

% \begin{rem}
~\label{rem:ComBoxSAT} The complexity of the evaluation of
SAT$(n,m)(y=y_{n-1}y_{n-2}\cdots $ $y_{1}$ $y_{0})$ could be considered to be $%
\mathbf{O}(1)$. Instead of using a cycle, it is plausible to
consider that \SSAT is a circuit of logical gates. This is
depicted in figure~\ref{fig:BoxSATnxm}. Hereafter, SAT$(n,m)$
correspond to a logic circuit of "and", "or" gates, and the
complexity of its evaluation is $\mathbf{O}(1)$.
% \end{rem}

\begin{proposition}
~\label{prop:BuildSolSSAT}

\SSAT \ has different row formulas, and $m \leq 2^{n}$. Any subset
of $\Sigma^n$ could be a solution for an appropriate \SSAT.
\begin{proof}

 $\emptyset$ is the solution of a blocked board., i.e.,  for any
SSAT$(n,m)$ with $m=2^n$.

For $m=2^n-1$, it is possible to build a SSAT$(n,m)$  with only $x$ as
the solution. The blocked numbers $[0,2^n-1]$ $\setminus$ $\{x,
\overline{x}\}$ and $x$ is translated and added to SSAT. By
construction, SSAT$(n,m)$$(x)=1.$

 For $f$ different solutions. Let $S$=$\{x_1,\ldots,x_f\}$ be the given solutions.
 Then the blocked numbers
$B$ = $\{y \in [0,2^n-1] \, | \, \forall x \in S_X, y \neq x
\text{ or } \overline{y} \neq x \}$, where $S_X$ $=$ $\{s \in X \,
| \,  \overline{s} \notin X \}.$ The numbers of $S_X$ and
$B\setminus X$
 are translated and added to the resulting SSAT.
\end{proof}
\end{proposition}

It is prohibitive to analyze more than one iteration SSAT's
formulas. For example, when $m\approx 2^{n},$ any strategy for
looking solving SSAT could have $m$ as a factor of the later
iterations.
% \end{rem}

\begin{proposition}
~\label{prop:EvalMatchFixedPoint} $y\in \Sigma ^{n}$ and $
y=y_{n-1}y_{n-2}\cdots y_{1}y_{0}.$ The following strategies of
resolution of SAT$(n,m)$ are equivalent.

\begin{enumerate}

\item The evaluation of SAT$(n,m)(y)$ as logic circuit.

\item~\label{stp:match} A matching procedure that consists
verifying that each $y_{i}$ match at least one digit $s_{i}^{k}\in
M_{n\times m},$ $\forall k=1,\ldots ,m$.

\end{enumerate}

\begin{proof}
SAT$(n,m)(y)=1$, it means that at least one variable of each row
is 1, i.e., each $y_i,$ $i=1,\ldots,n$ for at least one bit, this
matches to 1 in $s^k_j$, $k=1,\ldots, m$.
\end{proof}
\end{proposition}

% \begin{rem}
The evaluation strategies are equivalent but the computational
cost is not. The strategy~\ref{stp:match} implies at least $m
\cdot n$ iterations. This is a case for using each step of a cycle
to analyze each variable in a row formulas or to count how many
times a Boolean variable is used.
% \end{rem}

\begin{proposition}
An equivalent formulation of \SSAT\ is to look for a binary number
$x^{\ast }$ from $0$ to $2^{n}-1.$

\begin{enumerate}
\item If $x^{\ast }\in M_{n\times m}$ and $\overline{x}^{\ast
}\notin M_{n\times m}$ then SAT$(n,m)(x^{\ast })=1.$

\item If $x^{\ast }\in M_{n\times m}$ and $\overline{x}^{\ast }\in
M_{n\times m} $ then SAT$(n,m)(x^{\ast })=0.$ If $m<2^{n}-1$ then
$\exists y^{\ast} \in [0, 2^{n}-1]$ with $\overline{y}^{\ast
}\notin M_{n\times m} $ and SAT$(n,m)(y^{\ast })=1.$

\item if 2), then $\exists$ SAT$(n,m+1)$ such that 1) is fulfill.
\end{enumerate}

\begin{proof}  \

\begin{enumerate}
\item When $x^{\ast }\in M_{n\times m}$ and $\overline{x}^{\ast
}\notin M_{n\times m}$, this means that the corresponding formula
of $x^{\ast }$ is not blocked and for each Boolean formula of
SAT$(n,m)(x^{\ast })$ at least one Boolean variable coincides with
one variable of $x^{\ast }.$  Therefore SAT$(n,m)(x^{\ast })=1.$

\item I have,  $m<2^{n}-1$, then $\exists y^{\ast } \in[0,
2^{n}-1]$ with $\overline{y}^{\ast }\notin M_{n\times m}.$
Therefore, \SSAT$(y^\ast)=1$.

\item Adding the corresponding formula of $y^{\ast }$ to
SAT$(n,m)$, a new SAT$(n,m+1)$ is obtained. By 1, the case is
proved.
\end{enumerate}
\end{proof}
\end{proposition}

\begin{proposition}
~\label{prop:NumbSolSSAT}

SSAT$(n,m)$ has different row formulas, and $m \leq 2^{n}$.

 The complexity to solve SSAT$(n,m)$ is $\mathbf{O}(1).$

\begin{proof}

With the knowledge that $m < 2^n $ the Boolean formulas of SSAT$(n,m)$ does not correspond to a blocked board.
    It has not
    solution when $m=2^n$ and the SSAT$(n,m)$'s rows are different,
     i.e., it is a blocked board.
\end{proof}
\end{proposition}

This approach allows  for verifying and getting a solution for any
\SSAT. By example, SAT$(6,4)$ corresponds to the set $M_{6\times
4}$:

\begin{equation*}
\begin{tabular}{|l|l|l|l|l|l|l|}
\hline
& $x_{5}=0$ & $x_{4}=0$ & $x_{3}=0$ & $x_{2}=0$ & $x_{1}=0$ & $x_{0}=0$ \\
\hline & $\overline{x}_{5}\vee $ & $\overline{x}_{4}\vee $ &
$\overline{x}_{3}\vee $
& $\overline{x}_{2}\vee $ & $\overline{x}_{1}\vee $ & $\overline{x}_{0})$ \\
\hline
$\wedge ($ & $\overline{x}_{5}\vee $ & $\overline{x}_{4}\vee $ & $\overline{x%
}_{3}\vee $ & $\overline{x}_{2}\vee $ & $\overline{x}_{1}\vee $ &
$x_{0})$
\\ \hline
$\wedge ($ & $x_{5}\vee $ & $x_{4}\vee $ & $x_{3}\vee $ & $x_{2}\vee $ & $%
x_{1}\vee $ & $\overline{x}_{0})$ \\ \hline
$\wedge ($ & $\overline{x}_{5}\vee $ & $x_{4}\vee $ & $x_{3}\vee $ & $%
\overline{x}_{2}\vee $ & $x_{1}\vee $ & $x_{0})$ \\ \hline
\end{tabular}%
\text{ \ }%
\end{equation*}

\begin{equation*}
\begin{tabular}{|l|l|l|l|l|l|}
\hline $x_{5}$ & $x_{4}$ & $x_{3}$ & $x_{2}$ & $x_{1}$ & $x_{0}$
\\ \hline $0$ & $0$ & $0$ & $0$ & $0$ & $0$ \\ \hline $0$ & $0$ &
$0$ & $0$ & $0$ & $1$ \\ \hline $1$ & $1$ & $1$ & $1$ & $1$ & $0$
\\ \hline $0$ & $1$ & $1$ & $0$ & $1$ & $1$ \\ \hline
\end{tabular}%
\text{.}
\end{equation*}

The first table depicts that SAT$(6,4)(y=000000)=1$. The second
table depicts the set $M_{6\times 4}$ as an array of binary
numbers. The assignation $y$ corresponds to first row of
$M_{6\times 4}.$ At least one digit of $y$ coincides with each
number of
M$_{n\times m}$, the Boolean formulas of SAT$(6,4).$ Finally, $y$ $=$ $%
000000$ can be interpreted as the satisfied assignment $x_{5}=0,$
$x_{4}=0,$ $x_{3}=0,$ $x_{2}=0,$ $x_{1}=0,$ and $x_{0}=0.$

\SSAT can be used as an array of $m$ indexed Boolean formulas. In
fact, the previous proposition gives an interpretation of the
\SSAT \ as a type fixed point problem. For convenience, without
 exploring the formulas the SAT, my strategy is to look each
formula, and to keep information in a Boolean array of the
formulas of SAT by its binary number as an index for the array. At
this point, the resolution \SSAT \  is equivalent to look for a
binary number $x$ such that \SSAT$\left( x\right) =1$. The
strategy is to use the binary number representation of the
formulas of \SSAT \  in M$_{n\times m}.$

SSAT as a function can be see as the function of a fixed point
method, however, a satisfactory assignation could not belong to
the binary translations of the SSAT's formulas. The advantage of
taking the candidates from translations of SSAT's formulas is that
for each failure, two numbers can be discarded (see
prop.~\ref{prop:SAT_TwoOne}).

Furthermore, the equivalent between  SSAT  with the alternative
formulation to determine if there is a binary string, which is not
blocked in binary translations of the SSAT's formulas point out
the lack of relationship between the rows of SSAT.

%%%%%%%%%%%%%%%%%%%%%%%%%%%%%%%%%%%%%%%%%%%%%%%%%%%%%%%%%%%%%%%%%%%%%%%%%%%%
%%%%%%%%%%%%%%%%%%%%%%%%%%%%%%%%%%%%%%%%%%%%%%%%%%%%%%%%%%%%%%%%%%%%%%%%%%%%
\section{Extreme SSAT Problem}~\label{sc:exSSATCon}
%%%%%%%%%%%%%%%%%%%%%%%%%%%%%%%%%%%%%%%%%%%%%%%%%%%%%%%%%%%%%%%%%%%%%%%%%%%%
%%%%%%%%%%%%%%%%%%%%%%%%%%%%%%%%%%%%%%%%%%%%%%%%%%%%%%%%%%%%%%%%%%%%%%%%%%%%
%%%%%%%%%%%%%%%%%%%%%%%%%%%%%%%%%%%%%%%%%%%%%%%%%%%%%%%%%%%%%%%%%%%%%%%%%%%%
%%%%%%%%%%%%%%%%%%%%%%%%%%%%%%%%%%%%%%%%%%%%%%%%%%%%%%%%%%%%%%%%%%%%%%%%%%%%

In section~\ref{sc:SAS_SSAT}, the prop.~\ref{prop:NumbSolSSAT}
depicts that if SSAT's information includes that its rows are
different then to answer is not a complex problem. In fact, SSAT's
formulas are not necessary to review. The number of rows $m$ and
the fact that the SSAT's rows are different imply the answer
without viewing inside the given problem SSAT.

Here, let us be critical, in order to build with precision an
extreme problem. The extreme SSAT includes the parameters $n$
(number of Boolean variables) and $m$ the number of SSAT's rows.
No information about the specific of SSAT's rows are given. But,
the extreme problem could be a SSAT problem with only one binary
string as solution or none, and it includes duplicate and disorder
SSAT's rows. The selection of the unique solution is arbitrary,
i.e., it could be any $s \in [0,2^n-1]$. Hereafter, $\mathcal{S} =
\left\{x \in [0,2^n-1] \, | \, \text{SSAT}(n,m)(x)=1\right\}.$

The next propositions, depicts the difficult for determining a
satisfactory assignation for an extreme SSAT.

\begin{proposition}~\label{prop:probsel}
Let $n$ be large, and SSAT$(n,m)$ an extreme problem, i.e.,
$|\mathcal{S}|$ $\leq 1$, and $m \gg 2^n$.

\begin{enumerate}

\item The probability for selecting a solution (\Prob$_{ss}(f)$)
after testing $f$ different candidates ($f<<2^n$) is $ \approx 1 /
2^{2n}$ (it is insignificant).

\item Given $C$ $\subset$ $[0,2^n-1]$ with a polinomial
cardinality, i.e., $|C|$ $=$ $n^k$, with a constant $k
>0.$ The probability that the solution belongs  $C$ (\Prob$_{s}(C)$) is insignificant,
and more and more insignificant when $n$ grows.

\item  Solving SSAT$(n,m)$ is not efficient.
\end{enumerate}

\begin{proof} \

Assuming that $|\mathcal{S}|=1$.
\begin{enumerate}

\item The probability \Prob$_{ss}(f)$ corresponds to product of
the probabilities for be selected and be the solution.  For the
inner approach (i.e., the $f$ candidates are from the translations
of the SSAT$(n,m)$'s rows) \Prob$_{ss}(f)$ $=$ $1/\left( 2^{n}-2f
\right) \cdot 1 /2^n \approx 1/2^{2n} \approx 0.$ For the outside
approach (i.e., the $f$ candidates are from the $[0,2^n-1]$ the
search space) \Prob$_{ss}(f)$ $=$ $1/\left( 2^{n}-f \right) \cdot
1 /2^n \approx 1 /2^{2n} \approx 0.$

\item \Prob$(C)$ $=$ $n^k / 2^n.$ Then \Prob$_s(C)$ $=$ $n^k /
2^n$ $\cdot$ $1 /2^n$, and $lim_{n \to \infty} K n^k / 2^{n}$
(L'H\^{o}\-pi\-tal's rule) $=$ $0^+,$ $K>0.$ For $n$ large, $2^n-Kn^k
\approx 2^n,$ and $Kn^k \ll 2^n.$ Moreover, for the inner
approach, \Prob$_{ss}(n^k)$ $=$ $1/\left( 2^{n}-2n^k \right) \cdot
1 /2^n \approx 1/2^{2n} \approx 0.$ For the outside approach,
\Prob$_{ss}(n^k)$ $=$ $1/\left( 2^{n}-n^k \right) \cdot 1 /2^n
\approx 1 /2^{2n} \approx 0.$

 \item In any approach, inner or outside,
many rows of SSAT$(n,m)$ have large probability to be blocked,
because there is only one solution. Then the probability after $f$
iterations remains $1/ 2^{2n} \approx 0$. It is almost impossible
to find the solution with $f$ small or a polinomial number of $n$.
\end{enumerate}

Assuming that $|\mathcal{S}|=0$. \Prob$_s$ $=$ $0.$

\begin{enumerate}

\item[1,2] For the inner approach and for the outside approach,
\Prob$_{ss}(f)$ $=$ $0.$

\item[3] It is equivalent $\mathcal{S}=\emptyset$
$\Leftrightarrow$ SSAT$(n,m)(x)$
 $=$ $0,$ $\forall x \in [0,2^n-1].$ This means that it is
 necessary to test all the numbers in $[0,2^n-1].$

\end{enumerate}

\end{proof}
\end{proposition}

\begin{figure}[tbp]
\centerline{\psfig{figure=\IMAGESPATH/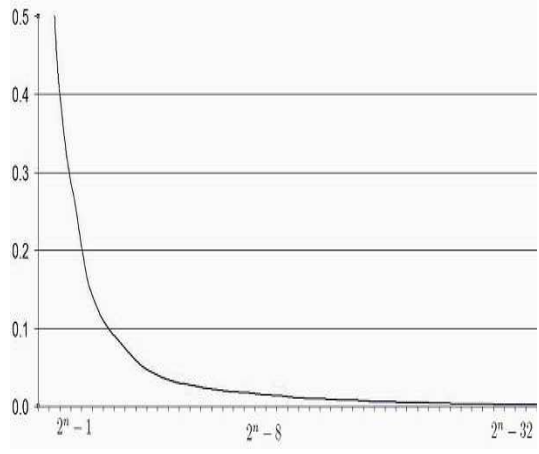,height=60mm,width=70mm
}}~ \caption{Behavior of the functions $P_e(t)$ and $P_i(t)$.}
\label{fig:probDecy}
\end{figure}

One important similarity between the extreme  SSAT as a numerical
problem (see prop.~\ref{prop:BuildSolSSAT}) for one or none
solution is the interpretation to guest such type of solution. It
is like a lottery but with the possibility that there is not
winner number. The exponential constant $2^n$ causes a  rapidly
decay as it depicted in fig.~\ref{fig:probDecy} where $t=2^n-1,
2^n-8, 2^n-32$.

The interpretation of taking the extreme SSAT as a circuit for an
electronic lottery behaves different when there is one winner
number than when there is none. It is probably to wait for long
time (it is an exponential waiting time) to get the winner number.
People accept the winner ticket $x^\ast,$ because a judge can show
in an electronic board the result SSAT$(x^\ast)$ $=$ $1$. It is
unlikely to get the winner ticket in short time, but most of the
people accept this case by testing the winner ticket. However, the
case when there is not winner number is rejected, because the long
time to wait to test all the numbers, and who can have the time,
and be the unconditional and unbiased witness to testify that
always the electronic board shows SSAT$(x)$ $=$ $0,$ $\forall
x\in[0,2^n-1]$. Both cases are similar, and they point out that
solving extreme SSAT takes an exponential time, no depending if a
group of person does a lottery or a computer performs an
algorithm.

\begin{proposition}
There is not an efficient algorithm for solving extreme \break
SSAT$(n,m)$.

\begin{proof}

If such algorithm exists then it is capable for solving in
polinomial time the equivalent number problem with one
winner number or none in contradiction to the exponential
time.
\end{proof}
\end{proposition}

%%%%%%%%%%%%%%%%%%%%%%%%%%%%%%%%%%%%%%%%%%%%%%%%%%%%%%%%%%%%%%%%%%%%%%%%%%%%
%%%%%%%%%%%%%%%%%%%%%%%%%%%%%%%%%%%%%%%%%%%%%%%%%%%%%%%%%%%%%%%%%%%%%%%%%%%%
\section{Algorithms for SAT} ~\label{sc:algthmsSSAT}
%%%%%%%%%%%%%%%%%%%%%%%%%%%%%%%%%%%%%%%%%%%%%%%%%%%%%%%%%%%%%%%%%%%%%%%%%%%%
%%%%%%%%%%%%%%%%%%%%%%%%%%%%%%%%%%%%%%%%%%%%%%%%%%%%%%%%%%%%%%%%%%%%%%%%%%%%
%%%%%%%%%%%%%%%%%%%%%%%%%%%%%%%%%%%%%%%%%%%%%%%%%%%%%%%%%%%%%%%%%%%%%%%%%%%%
%%%%%%%%%%%%%%%%%%%%%%%%%%%%%%%%%%%%%%%%%%%%%%%%%%%%%%%%%%%%%%%%%%%%%%%%%%%%

The previous sections depict characteristics and  properties of
SSAT. The complexity for solving any SSAT$(n,m)$ needs at least
one carefully review of the SSAT's rows, i.e., its complexity is
related to the numbers $m$, and any algorithm for solving SSAT
could have $m$ has a factor related to its complexity. If it uses
also the columns for substituting and simplifying by algebra the
factor grows $m \cdot n$ at least. Also, the ordering and
discarding repeated rows increased the complexity by
$m\text{log}_2(m)$. The properties depicted in
section~\ref{sc:SAS_SSAT} indicate two source of data for solving
SSAT$(n,m)$, 1) its $m$ rows or 2) the search space of all
possible Boolean values for its variables ($\Sigma^n$). The second
is large and $m$ could be large also. Therefore, the efficient
type of algorithms for solving SSAT must be doing in one way
without cycles, and with the constraint that the total iterations
must be related to $m < 2^n$, or $2^{n-1}$, or $2^n$. This is
because the fixed point approach or inside search(taking
candidates from the translation SSAT's formulas) and the outside
approach or probabilistic approach (taking candidates from the
search space $[0,2^n-1].$

It is necessary to be sceptical and impartial, in order to accept
the answer from a computer´s algorithm or a person. No matters if
$m$ is huge or SSAT is an extreme problem, without a proof or a
clearly explication, I reject to accept such answer. This impose
another characteristic for the algorithms for solving SSAT, they
must provide a witness or something to corroborate that SSAT has
been solved without objections.

A very simple algorithm to determine if SSAT has solution is
in~\cite{arXiv:Barron2015b}. The algorithm is presented to solve
SSAT by using the equivalent numerical formulation, more precisely
for  building an unsatisfactory board in a the table $T$.

\begin{algorithm}
~\label{alg:SATBoard_1} \textbf{Input:} \SSAT.

\textbf{Output:} The answer if \SSAT \ has solution or not. $T$ is
an unsatisfactory board when \SSAT\ has not solution.

\textbf{Variables in memory}: $T[0:2^{n}-1]$=$-1$: array of binary
integer; $address$: integer; $ct=0$: integer; $k$: binary integer.

\begin{enumerate}

\item  \textbf{if} $m < 2^n$ \textbf{then}

\item \hspace{0.5cm} \textbf{output:} "SSAT$(n,m)$ has a solution,

\hspace{0.5cm} its formulas  do not cover $\Sigma^n.$";

\item \textbf{end if}

\item \textbf{while not end(}\SSAT \textbf{)}

\item \hspace{0.5cm} $k=b_{n-1}b_{n-2}\ldots b_{0}$=
\textbf{Translate to binary formula} (\SSAT);

\item \hspace{0.5cm} \textbf{if }$k.[b_{n-1}]$ \textbf{equal} $0$
\textbf{then}

\item \qquad \qquad $address=2\ast k.[b_{n-2}\ldots b_{0}];$

\item \qquad \textbf{else}

\item \qquad \qquad $address=2\ast (2^{n-1}-k.[b_{n-2}\ldots
b_{0}])-1;$

\item \hspace{0.5cm} \hspace{0.5cm} \textbf{end if}

\item \hspace{0.5cm} \textbf{if} T[$address$] \textbf{equal} $-1$
\textbf{then}

\item \hspace{0.5cm}\hspace{0.5cm} $ct=ct+1$;

\item \hspace{0.5cm}\hspace{0.5cm} $T[address]=k;$

\item \hspace{0.5cm} \textbf{end if}

\item \hspace{0.5cm}\textbf{if} $ct$ \textbf{equal} $2^{n}$
\textbf{then}

\item \hspace{0.5cm} \hspace{0.5cm} \textbf{output:} "There is not
solution for \SSAT.

\hspace{0.5cm}   It has $2^n$ different formulas.";

\item \hspace{0.5cm} \hspace{0.5cm} \textbf{stop}

\item \hspace{0.5cm} \textbf{end if} \item \textbf{end while}
\item \textbf{output:} "\SSAT \ has a solution,

\hspace{0.5cm}  its formulas  do not cover $\Sigma^n.$";

\end{enumerate}
\end{algorithm}

The previous algorithm is quite simple. It does not require to
evaluate SSAT. The output has an equivalent formulation of the
input SSAT, as a table of un unsatisfactory board, it writes
"There is not solution for \SSAT". On the other hand, the
algorithm writes "\SSAT \ has a solution", without any additional
information, or witness.

It is reasonable to ask, do i accept the result of the previous
algorithm?. The answers is "yes" but after carefully reviewing and
verifying the correctness of the algorithm. If the answer of the
algorithm is forgotten, it is possible to recall the answer from
the table $T$, but it is not cheap. It is necessary to review in
order to determine if there is a binary number without its
complement or if all binary numbers are follow by its complement.
In the former case, SSAT has a solution, in the second no. The
objection is that the verification using $T$ after running the
algorithm is quite expensive.

Using the property of evaluating SSAT as circuit, the previous
algorithm is modified to the next algorithm.

\begin{algorithm}
~\label{alg:SATModBoard_1} \textbf{Input:} \SSAT.

\textbf{Output:} An unsatisfactory board T when \SSAT \ has not a
solution. A satisfactory assignation $k$ when \SSAT \ has a
solution.

\textbf{Variables in memory}: $T[0:2^{n}-1]$=$-1$: array of binary
integer; $address$: integer; $ct=0$: integer; $k$: binary integer.

\begin{enumerate}

\item  \textbf{if} $m < 2^n$ \textbf{then}

\item \hspace{0.5cm} \textbf{output:} "SSAT$(n,m)$ has a solution,

\hspace{0.5cm}  its formulas  do not cover $\Sigma^n.$";

\item \textbf{end if}

\item \textbf{while not end(}\SSAT \textbf{)}

\item \hspace{0.5cm} $k=b_{n-1}b_{n-2}\ldots b_{0}$=
\textbf{Translate to binary formula} (\SSAT);

\item  \hspace{0.5cm} \textbf{if} SSAT$(n,m)$($k$) \textbf{equal}
1 \textbf{then}

\item \hspace{0.5cm} \hspace{0.5cm} \textbf{output:} "$k$ is a
solution for SSAT$(n,m).$";

\item \hspace{0.5cm} \hspace{0.5cm}  \textbf{stop};

\item \hspace{0.5cm}  \textbf{end if};

\item \hspace{0.5cm} \textbf{if }$k.[b_{n-1}]$ \textbf{equal} $1$
\textbf{then}

\item \hspace{0.5cm} \hspace{0.5cm}  $k=\overline{k}$;

\item \hspace{0.5cm}  \textbf{end if};

\item \hspace{0.5cm} $address=2\ast k.[b_{n-2}\ldots b_{0}];$

\item \hspace{0.5cm} \textbf{if} T[$address$] \textbf{equal} $-1$
\textbf{then}

\item \hspace{0.5cm}\hspace{0.5cm} $T[address]=k;$

\item  \hspace{0.5cm}\hspace{0.5cm} $address=2\ast
(2^{n-1}-\overline{k}.[b_{n-2}\ldots b_{0}])-1;$

\item \hspace{0.5cm}\hspace{0.5cm} $T[address]=\overline{k};$

\item \hspace{0.5cm}\hspace{0.5cm} $ct=ct+2$;

\item \hspace{0.5cm} \textbf{end if}

\item \hspace{0.5cm}\textbf{if} $ct$ \textbf{equal} $2^{n}$
\textbf{then}

\item \hspace{0.5cm} \hspace{0.5cm} \textbf{output:} "There is not
solution for \SSAT.

\hspace{0.5cm}   It has $2^n$ different formulas.";

\item \hspace{0.5cm} \hspace{0.5cm} \textbf{stop}

\item \hspace{0.5cm} \textbf{end if}

\item \textbf{end while}

\item \textbf{for} $k=0$ to $2^{n}-1$ \textbf{do}

\item \hspace{0.5cm} \textbf{if} T[$k$] \textbf{equal} $-1$
\textbf{then}

 \item \hspace{0.5cm} \hspace{0.5cm} \textbf{output:}
 "$k$ is a solution of \SSAT.";

\item \hspace{0.5cm} \hspace{0.5cm}  \textbf{stop};

\item \hspace{0.5cm}  \textbf{end if};

\item \textbf{for};

\end{enumerate}
\end{algorithm}

The previous algorithm solves the problem and it provides two type
of witness: 1) an unsatisfactory board $T$ when there is no
solution, and 2) the satisfactory assignation $k$ when there is a
solution. It exploits the properties of \SSAT\ as a circuit, the
inside search (i.e., the candidates come from the SSAT's
formulas). Each failure eliminates two binary numbers, therefore
the table $T$ is building faster than the
algorithm~\ref{alg:SATBoard_1}. The algorithm does not use a
double linked list as the algorithms 2 and 3
in~\cite{arXiv:Barron2015b}. The drawback of this algorithm are
the last lines. Here, the satisfactory assignation is founded but
it is expensive with more the $2^n$ iterations. This could be
changed by using a double linked list as in algorithms 2 and 3
in~\cite{arXiv:Barron2015b}, this requires a lot of memory. The
difference between them is that the former stopped with one
satisfactory assignation and the second stopped after build
$\mathcal{S}$.

The algorithms 3 and 4 in~\cite{arXiv:Barron2015b} are building
using deterministic and probabilistic approach. They provides
different type of witness to corroborate when SSAT has solution or
not. The former gives a double linked list with the elements of
$\mathcal{S}$ and the other gives a Boolean table $T$ where the
elements of $\mathcal{S}$ correspond to $i\in[0,2^n-1]$ such that
$T[i]=0$.

 The situation for solving SSAT$(n,m)$ is subtle.
Its number of rows could be exponential, but for any SSAT$n,m)$,
there are no more than $2^n$ different rows, then $m \gg 2^n$
means duplicate rows. It is possible to consider duplicate rows
but this is not so important as to determine at least one solution
in $\Sigma^n$. The search space $\Sigma^n$ corresponds to a
regular expression and it is easy to build by a finite
deterministic automata (Kleene's Theorem) but in order. However,
to test the binary numbers in order is not adequate. For $m$ very
large any source of binary number as candidates must be random and
its construction be cheap. The next algorithm generates a random
permutation the numbers from 0 to $Mi$.

\begin{algorithm}~\label{alg:ParraBinNum} \textbf{Input:} $T[0: Mi]=[0:Mi]$.

\textbf{Output:} $T[0: Mi]$ contains a permutation of the numbers
from $0$ to $Mi$.

\textbf{Variables in memory}: $i=0$: integer; $rdm, a$=0 :
integer;

\begin{enumerate}

\item \textbf{for} i:=0 to $Mi-1$

\item \hspace{0.5cm} \textbf{if} $T[i]$ \textbf{equals} $i$
\textbf{then}

\item \hspace{0.5cm} \hspace{0.5cm} \textbf{select uniform
randomly} $rdm \in [i+1, Mi]$;

\item  \hspace{0.5cm} \hspace{0.5cm} $a$ $=$ $T[rdm]$;

\item \hspace{0.5cm} \hspace{0.5cm} $T[rdm]$ $=$ $T[i]$;

\item  \hspace{0.5cm} \hspace{0.5cm} $T[i]$ $=$ $a$;

\item \hspace{0.5cm} \textbf{end if}

\item \textbf{end for}

\item \textbf{stop}

\end{enumerate}
\end{algorithm}

An important property of this algorithm is that it builds a
permutation of the numbers $0$ to $Mi.$ None index coincide with
the numbers in order.

Let floor() be a function, it returns the smallest integer less
than or equal to a given number. Let rand() be a function that it
returns a random real number in $(0,1)$. The line \textbf{Select
uniform randomly} $rdm \in [0, Mi-1]$; could be implemented $k$
$=$ floor($r$ $\cdot$ $Mi$), where $r=rand()$, and $Mi>0$,
integer. Then $0$ $\leq$ $k$ $\leq Mi-1$. In similar way,
\textbf{Select uniform randomly} $rdm \in [i+1, Mi]$; could be
implemented as $k$ $=$ floor($r$ $\cdot$ $(Mi-i+1.5)$) + $(i+1)$.

The previous algorithm, is an alternative to change the line 4 in
the probabilistic algorithm 4 in~\cite{arXiv:Barron2015b}:

4. \hspace{0.5cm} \textbf{select uniform randomly} $k \in
[0,2^{n}-1] \setminus \{ i \, |\,  T[i] =1\}$;

Using the approach of the algorithm~\ref{alg:ParraBinNum}, the
next algorithm solves SSAT$(n,m)$ in straight forward using an
outside approach. Here, each candidates is a random selection from
$[0,2^{n-1}].$

\begin{algorithm}~\label{alg:SAT_Perm} \textbf{Input:} n, SSAT$(n,m)$.

\textbf{Output:} $rdm$, such that SSAT$(n,m)$$(rdm)=1$ or SSAT has
not solution.

\textbf{Variables in memory}: $T[0: 2^{n-1}-1]$=$[0: 2^{n-1}-1]$:
integer; $Mi$=$2^{n-1}-1$: integer; $rdm, a$: integer.

\begin{enumerate}

\item  \textbf{if} $m < 2^n$ \textbf{then}

\item \hspace{0.5cm} \textbf{output:} "SSAT$(n,m)$ has a solution,

\hspace{0.5cm}  its formulas  do not cover $\Sigma^n.$";

\item \textbf{end if}

\item \hspace{0.5cm} \textbf{if} $T[i]$ \textbf{equals} $i$
\textbf{then}

\item \textbf{for} i:=0 to $Mi-1$

\item \hspace{0.5cm} \textbf{if} $T[i]$ \textbf{equals} $i$
\textbf{then}

\hspace{0.5cm} // \textbf{select uniform randomly} $rdm \in [i+1,
Mi]$;

\item \hspace{0.5cm} \hspace{0.5cm} $rdm$ $=$ floor($rand()$
$\cdot$ $(Mi-i+1.5)$) $+$ $(i+1)$;

\item \hspace{0.5cm} \hspace{0.5cm} $a$ $=$ $T[rdm]$;

\item \hspace{0.5cm} \hspace{0.5cm} $T[rdm]$ $=$ $T[i]$;

\item  \hspace{0.5cm} \hspace{0.5cm} $T[i]$ $=$ $a$;

\item  \hspace{0.5cm} \textbf{end if}

\item \hspace{0.5cm} $rdm$ $=$ $0T[i]$;

 \item \hspace{0.5cm} \textbf{if}
SSAT$(n,m)$($rdm$) \textbf{ equals} 0  \textbf{and}

\hspace{0.9cm} SSAT$(n,m)$($\overline{rdm}$)   \textbf{ equals} 0
 \textbf{then}

\item \hspace{0.5cm} \hspace{0.5cm} \textbf{ continue}

\item  \hspace{0.5cm} \textbf{end if}

 \item \hspace{0.5cm} \textbf{if}
SSAT$(n,m)$($rdm$) \textbf{equals} 1 \textbf{then}

\item \hspace{0.5cm} \hspace{0.5cm} \textbf{output:} "$rdm$ is a
solution for SSAT$(n,m)$.";

\item \hspace{0.5cm} \hspace{0.5cm} \textbf{stop};

 \item \hspace{0.5cm} \textbf{else}

\item \hspace{0.5cm} \hspace{0.5cm} \textbf{output:}
"$\overline{rdm}$ is a solution for SSAT$(n,m)$.";

\item \hspace{0.5cm} \hspace{0.5cm} \textbf{stop};

\item  \hspace{0.5cm} \textbf{end if}

\item \textbf{end for}

\item  $rdm$ = $0T[Mi]$;

 \item  \textbf{if}
SSAT$(n,m)$($rdm$) \textbf{equal} 1 \textbf{then}

\item \hspace{0.5cm} \textbf{output:} "$rdm$ is a solution for
SSAT$(n,m)$.";

\item \hspace{0.5cm} \textbf{stop};

 \item \textbf{end if}

 \item \textbf{if}
SSAT$(n,m)$($\overline{rdm}$) \textbf{equal} 1 \textbf{then}

\item  \hspace{0.5cm} \textbf{output:} "$\overline{rdm}$ is a
solution for SSAT$(n,m)$.";

\item  \hspace{0.5cm} \textbf{stop};

\item  \textbf{end if}

\item \textbf{output:} "There is not solution for SSAT$(n,m)$,

SSAT$(n,m)(x)=0,$ $\forall x \in$ $[0,2^{n-1}]$.";

\item \textbf{stop};
\end{enumerate}
\end{algorithm}

The limit of the iterations to reach the answer is
$Mi+1=(2^{n-1}-1)+1=2^{n-1}$. Therefore, the complexity of the
previous algorithm is
$\mathbf{O}\left(Mi\right)$=$\mathbf{O}\left( 2^{n-1} \right)$. No
matters if the rows of SSAT$(n,m)$ are duplicates or disordered or
$m \gg 2^n$. The upper bound of the iterations is $2^{n-1}$ and
the search space is $[0,2^n-1]$ because a value
$x\in[0,2^{n-1}-1]$ is used to build $rdm=0x \in [0, 2^n-1]$ and
$\overline{rdm} \in [0, 2^n-1]$
 are tested in the same iteration.

%%%%%%%%%%%%%%%%%%%%%%%%%%%%%%%%%%%%%%%%%%%%%%%%%%%%%%%%%%%%%%%%%%%%%%%%%%%%
%%%%%%%%%%%%%%%%%%%%%%%%%%%%%%%%%%%%%%%%%%%%%%%%%%%%%%%%%%%%%%%%%%%%%%%%%%%%
\section{Complexity for SSAT}
~\label{sc:compleForSSAT}
%%%%%%%%%%%%%%%%%%%%%%%%%%%%%%%%%%%%%%%%%%%%%%%%%%%%%%%%%%%%%%%%%%%%%%%%%%%%
%%%%%%%%%%%%%%%%%%%%%%%%%%%%%%%%%%%%%%%%%%%%%%%%%%%%%%%%%%%%%%%%%%%%%%%%%%%%
%%%%%%%%%%%%%%%%%%%%%%%%%%%%%%%%%%%%%%%%%%%%%%%%%%%%%%%%%%%%%%%%%%%%%%%%%%%%
%%%%%%%%%%%%%%%%%%%%%%%%%%%%%%%%%%%%%%%%%%%%%%%%%%%%%%%%%%%%%%%%%%%%%%%%%%%%

The prop.~\ref{prop:BuildSolSSAT} depicts the complexity of
solving SSAT and how to build a SSAT with some given set of
solutions. By example, the following SSAT$(3,7)$ has one solution
$x_2=0$, $x_1=1$, and $x_0=1$:
\begin{equation*}
\begin{tabular}{llllccc}
& & & & & $\Sigma^3$ & $[0,7]$ \\
 &  $\overline{x}_{2}\vee $ & $\overline{x}_{1}\vee $ & $\overline{x}_{0})$ & & 000 & 0\\

$\wedge ($  & $\overline{x}_{2}\vee $ & $\overline{x}_{1}\vee $ &
$x_{0})$ & & 001 & 1
\\

$\wedge ($  & $\overline{x}_{2}\vee $ & $%
x_{1}\vee $ & $\overline{x}_{0})$ & &010 & 2\\

 $\wedge ($  & $%
\overline{x}_{2}\vee $ & $x_{1}\vee $ & $x_{0})$  & & 011 & 3 \\

$\wedge ($  & $x_2\vee $ & $\overline{x}_{1}\vee $ & $x_{0})$ & & 101 & 5\\
$\wedge ($  & $x_2\vee $ & $x_{1}\vee $ & $\overline{x}_{0})$ & & 110 & 6\\
$\wedge ($  & $x_2\vee $ & $x_{1}\vee $ & $x_{0})$  & & 111 & 7\\
\end{tabular}%
\text{ \ }%
\end{equation*}

By construction, the unique solution is the binary string of $3$.
It corresponds to the translation $( \overline{x}_{2}\vee
x_{1}\vee x_{0})$. It satisfies SSAT$(3,7)$, as the assignation
$x_2=0$, $x_1=1$, and $x_0=1$. It is not blocked by $100$, which
corresponds to the missing formula $(x_2 \vee \overline{x}_{1}\vee
\overline{x}_{0})$ (The complement of the formula $3$). The other
numbers $0,1,2$ are blocked by $5,6,7$.

\begin{proposition}~\label{prop:SSATBynSearch}
Let  SSAT$(n,2^n-1)$ be a problem with only one solution and its
rows in ascendent order. Then the complexity by a binary search to
determine the unique solution is
$\mathbf{O}\left(\log_2(2^n-1)\right)$ $\approx$
$\mathbf{O}\left(n \right)$.

\begin{proof}
Without loss of generality the rows can be as the previous example
SSAT$(3,7)$ in a table with indexes from $[0,2^n-2]$.

 The following algorithm determines the unique solution:

\begin{algorithm}
~\label{alg:SATBynSearch} \textbf{Input:} \SSAT\ with only one
solution and its rows in ascending order.

\textbf{Output:} The unique  satisfactory assignation $k$.

\textbf{Variables in memory}: $T[0:2^{n}-2]$ = (Translated
SSAT'rows) : array of binary integer; $l_i,r_i,m_i$: integer.

\begin{enumerate}

\item \textbf{if} $T[0]$ \textbf{is not equals} $0$ \textbf{then}

\item  \hspace{0.5cm} \textbf{output:} "$0$ is the solution.";

\item  \hspace{0.5cm} \textbf{stop};

\item \textbf{end if}

\item \textbf{if} $T[2^n-2]$ \textbf{is equals} $2^n-1$
\textbf{then}

\item  \hspace{0.5cm} \textbf{output:} "$2^n-1$ is the solution.";

\item  \hspace{0.5cm} \textbf{stop};

\item \textbf{end if}

\item $l_i=0$;

\item $r_i=2^n-2$;

\item \textbf{while} $((r_i - l_i) > 1)$ \textbf{do}.

\item  \hspace{0.5cm} $m_i = (l_i + r_i)/2. $

\item  \hspace{0.5cm} \textbf{if} $T[m_i]$ \textbf{is equals}
$m_i$ \textbf{then}

\item  \hspace{0.5cm} \hspace{0.5cm} $l_i=m_i$;

\item  \hspace{0.5cm} \textbf{otherwise}

\item  \hspace{0.5cm} \hspace{0.5cm} $r_i=m_i$.

\item  \hspace{0.5cm} \textbf{end if}

\item \textbf{end while}

\item \textbf{output:} "$l_i+1$ is the solution.";

\item \textbf{stop};

\end{enumerate}
\end{algorithm}

\end{proof}
\end{proposition}

The previous proposition is based in the numerical translation of
SSAT. The drawback of the previous binary search is that it only
applies for solving special SSAT$(2,2^n-1)$ with different rows
and in ascending order. When SSAT$(2,2^n-1)$'s rows are in
disorder, the cost of sorting  includes $\textbf{O}(2^n-1)$ by
using the Address Calculation Sorting (R. Singleton,
1956)~\cite{Agt:pena}. It has lineal complexity and is the less
expensive sorting to my knowledge. In this case the complexity to
determine the unique solution is $\textbf{O}(2^n).$

On the other hand, the no solution case has complexity
$\textbf{O}(1)$, knowing that SSAT$(n,2^n)$  has different rows,
there is nothing to look for. But again, to know that
SSAT$(n,2^n)$ has different rows, it has the cost of at least
$\textbf{O}(2^{n-1})$ by verifying at least one time the
SSAT$(n,2^n)$'s rows by using the algorithm 3 in
~\cite{arXiv:Barron2015b}.

The extreme SSAT problem is designed to test how difficult is to
determine one or none solution without more knowledge than $n$ the
number of variables, and $m$ the number of rows.  It is extreme
because $m \gg 2^n$ could be huge. This implies that SSAT'rows are
repeated, and the inner approach is not convenient. It could take
more than $m \gg 2^n$ iterations. Also, it does not help to know
that SSAT could have one or none solution. As it is mentioned
before, any algorithm must to solve SSAT without loops.

The algorithms 1,2, and 3 in~\cite{arXiv:Barron2015b} are based in
the inner or fixed point approach, therefore solving the extreme
SSAT could takes more than $2^n$ iterations ($m \gg 2^n$ is huge).
They  behave not stable for the extreme SSAT. The number of
iterations is quite wide depending of $m \gg 2^n$. With many
SSAT's rows repeated the inner approach or fixed point type method
has not advantage using the elimination of two candidates for
solving the extreme SSAT, it has to review the SSAT's rows but
duplicates rows do not provide information for knowing is the
solution or not solution is reached. It has the lower bound
$2^{n-1}$ for special SSAT$(2^n,2^n-1)$ because, it eliminates $k$
and $\overline{k}$ when $k$ cames from the translation of the
SSAT's rows. But depending if the SSAT's rows are duplicates and
disorder, it could behave quite different and makes a huge number
of iterations ($\gg 2^n$) for an extreme SSAT. By example, if SSAT
has the same row $2^n$ times at the beginning, after $2^n$
iterations the algorithm is far away for solving SSAT. This
phenomena does not happen with the outside approach, after $2^n$
iterations the solution is reached.

The algorithm~\ref{alg:SAT_Perm} is based in the outside approach.
It uses a random search in $[0,2^n-1]$ by creating two candidates
from $[0,2^{n-1}]$. The candidates are $0x$ and $\overline{0x}$,
$x\in [0,2^{n-1}].$ The pay off is an stable behavior, no matters
the extreme SSAT. Each candidate provides information that slowly
and consistently, it reduces the distance to the solution. When
there is no solution, this algorithm always takes $2^{n-1}$
iterations and it performs less than $2^{n-1}$ when there is one
solution. The algorithm takes advantage of the evaluation of SSAT
as a logic function in a circuit (see fig.~\ref{fig:BoxSATnxm})but
it can not use the inner approach´s property for eliminating two
candidates in each failure test but it tests two candidates at
same time.

The narrow behavior of the outside approach is the size of the
search space $[0,2^{n-1}]$. The wide behavior of the inner
approach is caused when $m$ $\gg 2^n$ and by the possibility for
testing  all SSAT's rows.

\begin{proposition}~\label{prop:alghtmsRange}
Let $n$ be large, and let SSAT be an extreme problem, i.e.,
$|\mathcal{S}|$ $\leq 1$. The algorithms 1, 2, and 3
in~\cite{arXiv:Barron2015b}, and algorithms~\ref{alg:SATBoard_1}
(inner approach) behaves wide, and the algorithms
4~\cite{arXiv:Barron2015b}, and algorithm~\ref{alg:SAT_Perm}
(probabilistic and outside approach) behave narrow.

\begin{proof} \
%%%%%%%%%%%%%%%%%%%%%%%%%%%%%%%%%%
The property depicted in prop.~\ref{prop:binNumBlock} relates $k$
and its complement, it allows to eliminate two numbers when the
candidate come from translation of a SSAT's formula. This is the
inner approach or fixed point type method. For the extreme SSAT,
any of the algorithms 1, 2, and 3  in~\cite{arXiv:Barron2015b},
and algorithms~\ref{alg:SATBoard_1} could iterates more than $2^n$
when the given SSAT's rows are repeated. In this case after $2^n$
iterations, it is possible to be far away of the solution. When
there is not solution, the number of iterations could be around $m
\gg 2^n$. It is a wide range of iterations from 1 to $m$ with $m$
$\gg$ $2^n$.

On the other hand, the algorithms 4~\cite{arXiv:Barron2015b}, and
algorithm~\ref{alg:SAT_Perm} (probabilistic and outside approach)
uses SSAT as function and they explores the search space
$[0,2^n-1]$ by creating two candidates from $[0,2^{n-1}].$ It
means that at most $2^{n-1}$ iterations are needed for solving any
SSAT, even in the case of an extreme SSAT with $m \gg 2^n.$

\end{proof}
\end{proposition}

\begin{proposition}
~\label{prop:SATVerifyNot} Given an extreme SAT$(n,m)$. It is not
possible to verify in polynomial time the solution of it.
\begin{proof}
This result follows from an extreme SSAT$(n,m)$, $m\gg 2^n.$ A
sceptical person or a computer program must matched the huge data
of SSAT$(n,m$) and the answer of the algorithms. He or it does not
execute any of the algorithms, they just receive the results. When
there is not solution, a table or an structure provide by the
algorithm means that $\mathcal{S}$ is empty. All the algorithms
here give an answer and a witness. It is simple to verify when
there is a solution $s^\ast$, SSAT$(n,m)(s^\ast)=1$. But, when the
answer is no solution, he or it has an equivalent formulation of
$\mathcal{S}=\emptyset$ or that the extreme SSAT$(n,m)$ is
equivalent to the special SSAT$(n,2^n)$ with different rows. The
corroboration can not consist in accepting the answer blindly
$\mathcal{S}=\emptyset$ or that the extreme SSAT$(n,m)$ is
equivalent to the special SSAT$(n,2^n)$. Also it is not sufficient
testing some candidates with SSAT$(n,m)$ but all. The
corroboration of the equivalence between extreme SSAT$(n,m)$ and
special SSAT$(n,2^n)$ needs at least $2^{n}$ iterations to match
their rows. Without executing a complete and carefully checking
and matching, the results of the algorithms themselves are not a
corroboration that the original extreme SSAT$(n,m)$ fulfill:
SSAT$(n,m)(x)$ $=$ $0,$ $\forall x \in [0,2^n-1]$ when there is
not solution.
\end{proof}
\end{proposition}

\begin{table}
  \centering
  \[
\begin{tabular}{|l|l|l|l|l|l|l|l|}
\hline
\multicolumn{2}{|l}{ $m\approx 2^{n}$} &
\multicolumn{3}{|l}{Existence} & \multicolumn{3}{|l|}{Construction} \\ \hline
\multicolumn{2}{|l}{SSAT$(n,m)$} & \multicolumn{3}{|l}{Test: $m-r\leq 2^{n}$}
& \multicolumn{3}{|l|}{$x\in $SSAT$(n,m)$} \\ \hline
rows & \vbox{\hbox{$r$ duplicate} \hbox{rows}} & min & avg & max & min & avg & max \\
\hline
$m=2$ & $0$ & $1$ & $1$ & $1$ & $1$ & $2$ & $3$ \\ \hline
$m<2^{n}$ & $0$ & $1$ & $1$ & $1$ & $1$ & ${m}/{2}$ & $m+1$ \\ \hline
$m=2^{n}-1$ & $0$ & $1$ & $1$ & $1$ & $1$ & $2^{n-1}$ & $2^{n}$ \\
\hline
\multicolumn{2}{|l}{$m=2^{n}$ (different rows)} & $1$ & $1$ & $1$ & $1$ & $1$
& $1$ \\ \hline
\multicolumn{8}{|l|}{$2^{n}$ (unknow rows, $2^{n}$ different rows) SSAT$(n,m)$
no solution} \\ \hline
$m=2^{n}$ & $0$ & $2^{n-1}$ & $2^{n-1}$ & $2^{n-1}$
& $2^{n-1}$ & $2^{n-1}$ & $2^{n-1}$ \\ \hline
$m=2^{n}+1$ & $1$ & $2^{n-1}$ & $2^{n-1}$ & $2^{n-1}$
& $2^{n-1}$ & $2^{n-1}$ & $2^{n-1}$ \\ \hline
$m=2^{n}+r$ & $r$ & $2^{n-1}$ & $2^{n-1}$ & $2^{n-1}$
& $2^{n-1}$ & $2^{n-1}$ & $2^{n-1}$ \\ \hline
\multicolumn{8}{|l|}{$2^{n}$ (unknow rows,$2^{n}-1$ different rows) SSAT$%
(n,m)$ unique solution} \\ \hline
$m=2^{n}-1+1$ & $1$ & $2$ & $2^{n-2}$ & $2^{n-1}$ & $1$ & $%
2^{n-1}$ & $2^{n-1}$ \\ \hline
$m=2^{n}-1+r$ & $r$ & $2$ & $2^{n-2}$ & $2^{n-1}$ & $1$ & $%
2^{n-1}$ & $2^{n-1}$ \\ \hline
\end{tabular}%
\]

  \caption{Behavior of algorithm~\ref{alg:SATModBoard_1} for the extreme SSAT}\label{tb:algInternal}
\end{table}

\begin{table}
  \centering
  \[
\begin{tabular}{|l|l|l|l|l|l|l|l|}
\hline \multicolumn{2}{|l}{$m\gg 2^{n}$} &
\multicolumn{3}{|l}{Existence} &
\multicolumn{3}{|l|}{Construction} \\ \hline
\multicolumn{2}{|l}{SSAT$(n,m)$} & \multicolumn{3}{|l}{ } &
\multicolumn{3}{|l|}{$x\in \left[ 0,2^{n-1}\right] $} \\ \hline
rows &\vbox{\hbox{$r$ duplicate} \hbox{rows}} & min & avg & max & min & avg & max \\
\hline \multicolumn{8}{|l|}{$2^{n}$ (unknow rows, $2^{n}$
different rows) SSAT$(n,m)$ no solution} \\ \hline $m=2^{n}+r$ &
$r$ & $2^{n-1}$ & $2^{n-1}$ & $2^{n-1}$ & $2^{n-1}$ & $2^{n-1}$ &
$2^{n-1}$
\\ \hline
\multicolumn{8}{|l|}{$2^{n}$ (unknow rows,$2^{n}-1$ different rows) SSAT$%
(n,m)$ unique solution} \\ \hline $m=2^{n}-1+r$ & $r$ & $1$ &
$2^{n-2}$ & $2^{n-1}$ & $1$ & $2^{n-2}$ & $2^{n-1}$ \\ \hline
\end{tabular}%
\]

  \caption{Behavior of algorithm~\ref{alg:SAT_Perm} for the extreme SSAT}\label{tb:algexternal}
\end{table}

The tables~\ref{tb:algInternal} and ~\ref{tb:algexternal} summarizes the complexity for solving the extreme SSAT. For solving extreme SSAT, the column existence depicts that the complexity is $\mathbf{O}(1)$ for almost all the cases but $m=2^n$ with unknown rows. This is because there is not a property for implying $\forall x \in [0,2^n-1],$ SSAT$(n,m)(x)=0$ but to verify that all SSAT's rows are different. For this case, the algorithms~\ref{alg:SATModBoard_1} and ~\ref{alg:SAT_Perm} prove that there is not solution after testing all possible candidates.

For more details see the end of the section~\ref{sc:exSSATCon}.
This means that there is no a shortcut for verifying $\mathcal{S}
= \emptyset$ for a given extreme SSAT$(n,m)$.

%%%%%%%%%%%%%%%%%%%%%%%%%%%%%%%%%%%%%%%%%%%%%%%%%%%%%%%%%%%%%%%%%%%%%%%%%%%%
%%%%%%%%%%%%%%%%%%%%%%%%%%%%%%%%%%%%%%%%%%%%%%%%%%%%%%%%%%%%%%%%%%%%%%%%%%%%
\section*{Conclusions and future work}
~\label{sc:conclusions and future work}
%%%%%%%%%%%%%%%%%%%%%%%%%%%%%%%%%%%%%%%%%%%%%%%%%%%%%%%%%%%%%%%%%%%%%%%%%%%%
%%%%%%%%%%%%%%%%%%%%%%%%%%%%%%%%%%%%%%%%%%%%%%%%%%%%%%%%%%%%%%%%%%%%%%%%%%%%
%%%%%%%%%%%%%%%%%%%%%%%%%%%%%%%%%%%%%%%%%%%%%%%%%%%%%%%%%%%%%%%%%%%%%%%%%%%%
%%%%%%%%%%%%%%%%%%%%%%%%%%%%%%%%%%%%%%%%%%%%%%%%%%%%%%%%%%%%%%%%%%%%%%%%%%%%

The results here does not change the SAT's complexity of the
article~\cite{arXiv:Barron2015b}.  It was interesting to analyze
with more details that SSAT problems and algorithms behaves quite
wide. Particularly, the inner or fixed point approach has not an
advantage for eliminating two candidates for extreme SSAT$(n,m)$
and it gives the wide behaviour. However, outside approach or
probabilistic approach behaves stable with the upper bound
$2^{n-1}.$

The outside approach and the evaluation of SSAT as a circuit
correspond to the probabilistic type of method allow to build the
stable algorithm~\ref{alg:SAT_Perm}. This algorithm is a more
detailed version of the probabilistic algorithm 4 of
~\cite{arXiv:Barron2015b}.

Moreover, for extreme SSAT $(n,m)$ with $m \approx 2^n$ the
complexity inside (alg.\ref{alg:SATModBoard_1}) is similar to the
outside (alg.\ref{alg:SAT_Perm}), i.e., $\mathbf{O}\left(
2^{n-1}\right).$

The main result is the impossibility to build an efficient
algorithm for solving the decision SSAT, i.e., for knowing if it
has a satisfactory assignation or not. The sceptical point of view
needs proof to confirm or deny an answer. The algorithms in this
paper always give some kind of witness or proof. When $m<2^n,$
there is a solution because the formulas of the given SSAT$(n,,m)$
do not cover the binary combination of the search space
$\Sigma^n.$ A satisfactory assignation when SSAT has solution is
sufficient. But, a message when there is not solution do not
substitute the detailed corroboration that SSAT$(n,m)$ has $2^n$
different formulas or that $\forall x \in[0,2^n-1]$,
SSAT$(n,m)(x)=0$ with $m \gg 2^n$. The lack of an easy test to
verify when there is not solution point out that there is not way
for verifying a solution in polynomial time.

Extreme SSAT states that in order to solve it, at least one review
of  its search space ($\Sigma^n$) is necessary. This is done by
splitting it into two spaces: $\mathbf{0}\Sigma^{n-1}$ and
$\mathbf{1}\Sigma^{n-1}$ in at most $2^{n-1}$ iterations. Finally,
this implies  $\mathbf{O}(\text{SSAT})=\mathbf{O}(2^{n-1})$
$\preceq$ $\mathbf{O}(\text{NP-Soft})$ $\preceq$
$\mathbf{O}(\text{NP-Hard}).$

%%%%%%%%%%%%%%%%%%%%%%%%%%%%%%%%%%%%%%%%%%%%%%%%%%%%%%%%%%%%%%%%%%%%%%%%%%%%
%%%%%%%%%%%%%%%%%%%%%%%%%%%%%%%%%%%%%%%%%%%%%%%%%%%%%%%%%%%%%%%%%%%%%%%%%%%%
% \bibliographystyle{abbrv}
% \bibliography{\BIBPATH/NPComplexity_v14}

\end{document}